\documentclass[10pt,journal,compsoc]{IEEEtran}

\usepackage[pdftex]{graphicx}
\usepackage{hyperref}
\usepackage{caption}
\usepackage{subcaption}

	\newcommand{\hN}[2]{\mathcal{N}(#1, #2)} 
	
	\usepackage{xcolor}

	\newcommand{\reffig}[1]{Fig.~\ref{#1}}
	
	\newcommand{\reftbl}[1]{Table~\ref{#1}}
	\newcommand{\refsec}[1]{Sec.~\ref{#1}}

	\usepackage{amsmath, amssymb,amsfonts,amsthm}

\begin{document}

\title{%
	Multi-layer network approach in modeling epidemics in an urban town
}%

\author{%
	Meliksah~Turker,
	Haluk~O.~Bingol
	\IEEEcompsocitemizethanks{
		\IEEEcompsocthanksitem M.~Turker and H.~O.~Bingol  are with 
		the Department of Computer Engineering,
		Bogazici University,
		Istanbul 34342,
		Turkey.
		
		Email: turkermeliksah@hotmail.com; bingol@boun.edu.tr.
	}%
}

\IEEEtitleabstractindextext{%
	\begin{abstract}
                The last three years have been an extraordinary time 
                with the Covid-19 pandemic killing millions, 
                affecting and distressing billions of people worldwide. 
                Authorities took various measures such as turning school and 
                work to remote and prohibiting social relations via curfews. 
                In order to mitigate the negative impact of the epidemics, 
                researchers tried to estimate the future of the pandemic for different scenarios, 
                using forecasting techniques and epidemics simulations on networks. 
                Intending to better represent the real-life 
                in an urban town in high resolution,
                we propose a novel multi-layer network model,
                where each layer corresponds to a different interaction that occurs daily,
                such as “household”, “work” or “school”.
                Our simulations indicate that locking down “friendship” layer 
                has the highest impact on slowing down epidemics.
                Hence, our contributions are twofold, 
                ﬁrst we propose a parametric network generator model; 
                second, we run SIR simulations on it and show the impact of layers.
	\end{abstract}
	
}

\maketitle

\section{Introduction}
The study of spread on networks provides insight about
how any diffusible such as disease, idea or gossip propagates on a network.
Understanding the process of diffusion, and the underlying network structure
allows taking actions to change the pace of diffusion,
such as declaring community lockdown to slow down an epidemic.

Most real-life networks are very complex and large to create an identical
twin to study.
This complexity leads researchers working on networks
to use either synthetic networks whose attributes are similar to real-life networks,
or domain specific and limited real-life networks.
Despite their similarity to real-world networks,
synthetic networks' ability to represent every real-world interaction is limited.
Moreover, domain specific real-world networks are either limited in size
or are also too specific to represent every aspect and interaction in life.

Recent Covid-19 pandemic has urged the research on
spread on networks~\cite{
	st2021social,%
	gosak2021community,%
	gosak2021endogenous,%
	markovivc2021socio,%
        yin2022impact,%
        wang2020epidemic,%
        sun2022diffusion,%
        fan2022epidemics}
and shown the need to model 
daily life interactions in a high resolution for accurate predictions and right interventions.

In order to support and enhance such studies,
we propose a parametric multi-layer network scheme to model the everyday interactions
between residents of a hypothetical urban town,
modeling individuals and interactions using undirected edge-weighted networks.

In our network, each individual in town is represented by a \emph{vertex},
and any physical interaction between two vertices that may spread a disease
is represented by an \emph{edge} with a \emph{weight} corresponding to transmission probability.
Since not all interactions have the same duration or intimacy,
different type of interactions pose different transmission probabilities;
hence they are represented by different edge weights.
For this reason, we adopt multi-layer network approach,
where each layer $\ell$ has its own $\beta_{\ell}$ edge weights.

In each layer of the network, we represent a fundamental relationship in daily life.
Following the bottom-up approach, the network is built from the most intimate 
and enduring relation to lesser ones. 
In total, the network consists of 7 layers,
namely;
household,
blue-collar workplace,
white-collar workplace,
school,
friendship,
service industry,
and finally, random encounters.

Moreover, vertices are placed onto locations on the network
and interact with their relative neighborhoods.
This approach of ``locality'' allows connecting vertices in a realistic way,
rather than randomly,
so that they make connections with other vertices
by going to work, school, shopping according to where they are,
just like the real-world.

The network is defined by two sets of parameters.
The first set of parameters define the static, broad structure
of the network such as network size, the ratio of workforce,
the ratio of vertices that go to school.
The second set of parameters define the distributions,
values such as household size,
number of students in a classroom,
number of friends a vertex has,
are sampled from.
This way, we obtain a diverse and non-homogeneous network,
rather than a lattice-like network e.g., where every vertex
would live in a house of 4 vertices and have 10 friends.
Both sets of parameters are obtained from the real-world whenever possible
and assumed plausible values otherwise.

We believe this parametric multi-layer network scheme
reflects what happens in an urban town in high resolution
and can be used to simulate and inspect different scenarios. 
The modularity of layers allows answering questions like
``How helpful is it to turn schools to remote?'',
``What would happen if both schools and white-collar jobs turned to remote?'', 
and
``What is the most impactful layer to slow down an epidemic?'' 
by means of inspecting
network attributes and running epidemic simulations.
We confirm the representative power of our model in two ways.
First, our SIR simulation results are aligned with 
the most recent research and the real-word data~\cite{%
	haug2020nature,%
	gosak2021community}.
Second, our networks' attributes are comparable to real-world networks,
shown in \refsec{sec:NetworkAttributes}
Moreover, even though the focus of this work is on epidemics due to
the recent Covid-19 outbreak, 
multi-layer network scheme can be used to study other fields
on network science such as idea and gossip propagation,
social behaviors, and game playing as well.

\section{Related Work}

Network science has been a widely studied area,
especially in the last decades with the increased amount of data.
Despite the increasing availability of data, real-world networks are often
limited in size, specific to certain domains and static 
since they are often snapshots taken in a particular moment.
Some examples to widely studied real-world networks are
Zachary's karate Club ($N = 34$)~\cite{zachary1977information},
Zambian tailor shop ($N = 39$)~\cite{kapferer1972strategy},
professional relationships among managers ($N = 21$)~\cite{krackhardt1987cognitive},
relationships among Lazega Law Firm partners ($N = 71$)~\cite{lazega2001collegial},
American football network ($N = 115$)~\cite{girvan2002community},
primary school contact network ($N = 236$)~\cite{stehle2011high},
where $N$ denotes the number of vertices, or network size.
Evidently, real-world networks are frequently two to three digits in size.

Models like Erdős–Rényi~\cite{erdosrenyi1959onrandom},
Watts-Strogatz~\cite{watts1998collective},
Barabási–Albert~\cite{barabasi1999emergence},
and random geometric graphs in hyperbolic spaces~\cite{boguna2010sustaining}
are able to dynamically and scalably generate networks whose attributes,
such as small diameter, short average path length,
strong clustering, and community structures are similar to real-world networks~\cite{%
	newman2006structure,%
	krioukov2010hyperbolic,%
	zuev2015emergence}.
Hence these models are commonly used on network science research.

Several works studied various types of interactions on networks, 
such as rumor and gossip propagation~\cite{zanette2002dynamics,
	lind2007spreading}, 
ideological opinion spread~\cite{volkening2020forecasting},
and finally physical, 
infectious relations that can spread disease~\cite{%
	may2001infection,%
	newman2002spread}. 
According to common approach in these works,
individuals or agents are represented as vertices of the network and
interactions or connections between vertices are represented
as edges.
Therefore, direct propagation of an idea or a disease between two vertices
is possible only if they are connected via an edge.

Many researchers~\cite{%
	moreno2002epidemic,%
	youssef2011epidemicsonweightedsir,%
	kamp2013epidemic,%
	wang2019multiplexawareness,%
	yin2022impact,%
	wang2020epidemic,%
	sun2022diffusion,%
	fan2022epidemics}
working on epidemics on networks
considered models like SI, SIS, and SIR
where S, I, and R stand for Susceptible, Infected and Recovered/Removed respectively.
In these models, an agent can be in one of the mentioned states at a time.
Initially, all agents in the population are in susceptible state.
Then some selected agents are infected with disease.
Susceptible agents that contact infected ones also become infected with
\emph{transmission probability} $\beta$.
Over time, infected agents either recover or die and get removed from the system,
with recovery probability $\gamma$.
From the perspective of research, both recovered and removed represent
the same state, hence they are used interchangeably and denoted by R.

It is convenient to use weighted networks~\cite{%
	gang2005epidemic,%
	youssef2011epidemicsonweightedsir,%
	kamp2013epidemic}
to represent the transmission probability
between two vertices as weighted edge.
This way, it becomes possible to model heterogeneous
interactions with various transmission probabilities.

Additional to epidemic spread on networks, 
vaccination on networks is also studied~\cite{%
	madar2004immunization,
	wang2016statistical},
where vaccination is represented by the removal of 
vaccinated vertices from the network~\cite{albert2000error},
hence rendering vaccinated agents immune, 
unable to get infected and spread the disease
onto other agents.

Multi-layer networks allow representing different types of interactions
in different sub-networks or layers.
References~\cite{
	marceau2011modeling,%
	sahneh2013generalized%
},
worked on epidemics on multi-layer networks, 
where there are multiple graphs or layers that share all vertices
but not all edges.
Each layer represents a different type of interaction
and agents interact through multiple layers.
In both of these papers, researchers use variations of SIS and SIR models
on top of two-layer synthetic networks,
where layers in the prior work are created by 
Molloy Reed algorithm~\cite{
	molloy1995critical%
}.
Similarly, Buono and her colleagues~\cite{%
	buono2014multiplex%
}
worked on epidemics on multi-layer complex networks,
representing various interactions through different layers,
which are also created by Molloy Reed algorithm.
In this work, vertices are partially overlapped;
therefore, not every vertex exists on every layer of the network.
Following these, Wang and his colleagues~\cite{%
	wang2019multiplexawareness,%
	wang2020co%
}
worked on awareness of epidemics in two-layer networks,
generated by Erdős–Rényi and Barabási–Albert models.
In these two-layer network schemes,
one layer propagates awareness about the disease
and the other layer propagates the disease.
Following the idea of separating the epidemic spread
and auxiliary means spread via different layers,
several recent works~\cite{%
	yin2022impact,%
	wang2020epidemic,%
	sun2022diffusion,%
	fan2022epidemics%
} 
studied the interaction of the two.
Types of auxiliary means in these works include disease information,
vaccination behavior, anti-vaccination propaganda, 
positive and negative preventive information,
and resources such as medical resources.
These works used Erdős–Rényi model to
construct the physical interaction network layer
where the disease spread occurs.

The impact of awareness on epidemics is studied 
in single layer networks~\cite{azizi2020epidemics} as well,
where both awareness and disease spread over the same network.
This work considered Erdős–Rényi~\cite{erdosrenyi1959onrandom},
Watts-Strogatz~\cite{watts1998collective} and
Barabási–Albert~\cite{barabasi1999emergence} models when creating networks.
Moreover, a recent work~\cite{su2022evolution}
studied evolutionary prisoner's dilemma on two-layer networks
where the researchers used both synthetic and real-world multi-layer networks.
Six real-world networks with network size ranging between
$N = 21$ to $N = 71$ are studied.
In order to obtain larger networks, synthetic networks are created using
Erdős–Rényi, Barabási–Albert and Goh-Kahng-Kim~\cite{goh2001universal} models.

Being aware of the difficulty of 
modeling millions of agents in individual level on a network,
a model is proposed to cluster vertices into groups
and use these clusters as the high level representation of the network to study epidemics
on large scale via approximation~\cite{prasse2021clustering}.
The method is applied to three real-world networks whose sizes vary
between $N = 64$ and $N = 236$.

Last but not least, a recent survey~\cite{jusup2022social},
touched upon the relevancy of social networks, internet search data,
and geographic location data to the epidemics, network science
and multi-layer networks.

\subsection{Covid-19}

Covid-19's emergence and lockdowns interrupted social life widely
and brought about more research onto epidemics and networks.
Despite their impact on slowing down epidemics,
lockdowns and quarantines caused 
mental and psychological issues on the societies~\cite{%
	sameer2020assessment}.
Naturally, several works studied the effectiveness of precautions and lockdowns.
A relatively early work conducted during the first phase of Covid-19 pandemics~\cite{%
	haug2020nature} 
inspected the outcome of precautions taken against Covid-19 
using statistical methods on evidential real-world data collected worldwide during 2020
and found that cancellation of small gatherings is the most impactful precaution
to slow down Covid-19.
Following, the impact of collective behavior to end epidemics
is studied~\cite{st2021social}, and it is suggested that 
blanket cancellation of events that are larger 
then a critical size can suddenly stop epidemics.
Gosak and his colleagues~\cite{gosak2021community}
studied whether lockdowns are effective
at slowing down epidemics, running SIR model on both 
synthetic random geometric graphs in hyperbolic spaces networks 
and real-life network of size $N = 58,000$,
obtained by merging phone location data
and two online social platforms.
The finding of this research is that
lockdowns alone have a low impact on slowing down epidemics.
Even though it is not a costly precaution like lockdown,
the impact of social distancing on epidemics is also studied
by Gosak and his colleagues again~\cite{gosak2021endogenous},
where the problem in question is formulated as
a game theory on networks.
In this work, SEIR simulation on synthetic
random geometric graphs in hyperbolic spaces network is run
and the results suggest that contact and social distancing is not static
as authorities and other researchers assume it to be,
and endogenous social distancing should be taken into account.

Two papers from 2021~\cite{%
	zhang2021beta013,%
	feng2021beta017}
worked on the transmission of Covid-19
and reported that the transmission rate for Covid-19 were 0.13 and 0.17, respectively.
Having these quantities is especially important in the perspective of
research of epidemics on networks with regards to Covid-19, since it allows
leveraging weighted networks with corresponding edge weights to represent
transmission probabilities.

The impact of vaccination and prioritization of vaccines in case of short supply
is also studied~\cite{markovivc2021socio},
where researchers inspected the efficiency of different vaccination strategies
to contain Covid-19, by running SEIRS simulation on synthetic
random geometric graphs in hyperbolic spaces network. 

A recent multi-layer network paper~\cite{
	bongiorno2022multi}
inspected the impact of vaccination
in France with focus on schools.
Despite being similar to our work,
the authors' version has significant differences
in network methodology as well as the aim of the research,
whereas our paper is a more generic network generator.

Throughout the literature of network science, it is observed that
researchers very frequently work on either limited in size and aspect, domain-specific
real-world networks, or synthetic random networks.
Our model aims to provide a scalable parametric network generator framework,
discussed in \refsec{sec:NetworkGenerator},
that represents several aspects of
physical and social interactions within an urban town
to endorse network science research.

The model proposed in this paper
leverages multi-layer weighted networks to build its layers.
Its multi-layer approach is similar to Buono's~\cite{buono2014multiplex}
in terms of partially overlapping vertices.
Moreover, edge weights represent the transmission probability~\cite{gang2005epidemic},
denoted by $\beta$.
We use a range of $\beta$ values to simulate various scenarios, 
including reported Covid-19 transmission rates~\cite{%
	zhang2021beta013,%
	feng2021beta017}.
Each layer has its own edge weights $\beta_{\ell}$ to represent
the various interactions with varying corresponding transmission probabilities.
Compared to the two-layer approaches in the literature~\cite{%
	buono2014multiplex,%
	wang2019multiplexawareness,%
	wang2020co,%
	wang2020epidemic,%
	sun2022diffusion,%
	fan2022epidemics%
	},
that use the auxiliary layer for the spread of information
and the other layer for the disease spreading;
our model consists of and spreads the disease on seven layers,
with each layer representing a fundamental interaction from daily life,
actuating different transmission rates.
To our knowledge, there is no similar work in the literature
in terms of
(i)~high resolution and representation power
offered by the number of layers,
(ii)~ability to represent diverse interactions in daily life through
different layers and transmission probabilities
(iii)~assignment of vertices to locations on ring lattice and locality of interactions via displacement.

\section{Network generator}
\label{sec:NetworkGenerator}

Assume a person is infected.
He or she can infect the household at home,
colleagues at work, friends during gatherings,
cashiers and other people in the market,
and his or her neighborhoods in the area.
These interactions differ in dynamics such as
number of connections,
the risk of transmission and the way they are created.
Multi-layer network structure is leveraged to model
different types of interactions separately and modularly.

\subsection{Concepts}

We begin by explaining the concepts on which we build our multi-layer network.
Note that the terms agent and vertex are used interchangeably.

\subsubsection{Interaction types and layers}

Close contact between a susceptible and infected creates a potential for disease
to spread from infected to susceptible.
This potential is implemented by the transmission probability.
However, not all real-life contacts are equally intimate, or of equal duration,
so they must be assigned transmission probabilities accordingly.

We construct a network
that is composed of seven layers.
Each layer represents a type of interaction
that can be associated with different levels of disease transmission.
Note that the vertices of the network are set at the beginning.
Therefore layers only add new edges between the vertices.
We define layers according to disease transmission probabilities and lockdown possibilities.

The first four layers are related to ``containers'' such as house or school.

\begin{itemize}
	
	\item 
	(L1)~\emph{Household layer} corresponds to interactions between households within a house. 
	House has the highest transmission probability among the containers 
	since interactions are more intimate and prolonged.
	Note that each agent should have exactly one house.
	
	\item
	(L2)~\emph{Blue-collar work layer} corresponds to workplace interactions between workers
	who still had to go to work even during the pandemic lockdown 
	because their jobs require them to be on site.
	Some examples to this type of work include work performed by workers of sectors such as logistics, 
	manufacturing, and couriers and cashiers of markets and suppliers, 
	as well as doctors and nurses.
	
	\item
	(L3)~\emph{White-collar work layer}, similar to blue-collar layer, 
	corresponds to interactions at work,
	except these interactions being occurring between people 
	who can work remotely via their computers such as
	office employees, software developers, text translators.
	There is no difference between blue-collar and white-collar workers normally,
	but these two-layers allow modeling lockdown and remote working.
	
	\item
	(L4)~\emph{School layer} corresponds to interactions between inhabitants of a school, 
	such as students, teachers, and other employees that work in it.
	
\end{itemize}

At each layer, there are a number of containers,
such as 
homes in house layer, 
classrooms in school layer and 
businesses in blue and white-collar layers.
Note that every agent is associated with one home.
A retired person is only associated with its home.
An agent may also be associated with a second container, 
such as a classroom if he or she is a student or teacher,
or to a business if he or she is a professional.
For assignment of agents to containers see \refsec{sec:AssignmentToContainers}.

Agents in a container are \emph{clique} connected,
i.e.,
all agents in the container are pairwise connected.
The number of agents in a container is called \emph{capacity}.
Therefore, 
it is possible that a student, that is infected by another student in the class, 
infects the households in her home.
And they go to work and infect their coworkers.

The remaining three layers are ``star'' connected.
In \emph{star} connection, a vertex $i$ at the center is connected to a group of vertices 
that are possibly not connected to each other.
The number of connections $i$ makes is called \emph{capacity}.
Interactions between workers in the service sector, 
and their customers,
any two friends,
and any random encounter
are represented by star connections.
Even though we model friendship as a star connection,
it is known that due to triadic closure,
friends of a person tend to be friends as well~\cite{easley2010networks}.
We leave that to the stochasticity of network generation.
\begin{itemize}
	
	\item 
	(L5)~\emph{Friendship layer} corresponds to interactions between friends, 
	such as a meeting between two friends.
	
	\item
	(L6)~\emph{Service industry layer} corresponds to interactions 
	between the employees of service industry,
	such as couriers and cashiers, and their customers.
	
	\item
	(L7)~\emph{Random encounters layer} corresponds to random interactions 
	between residents of a town
	that take place while shopping, in a restaurant or cafe, traveling or simply walking by on the street.
	
\end{itemize}

\begin{figure}[!tbp]
	\centering 
	\includegraphics[width=\columnwidth]%
        		{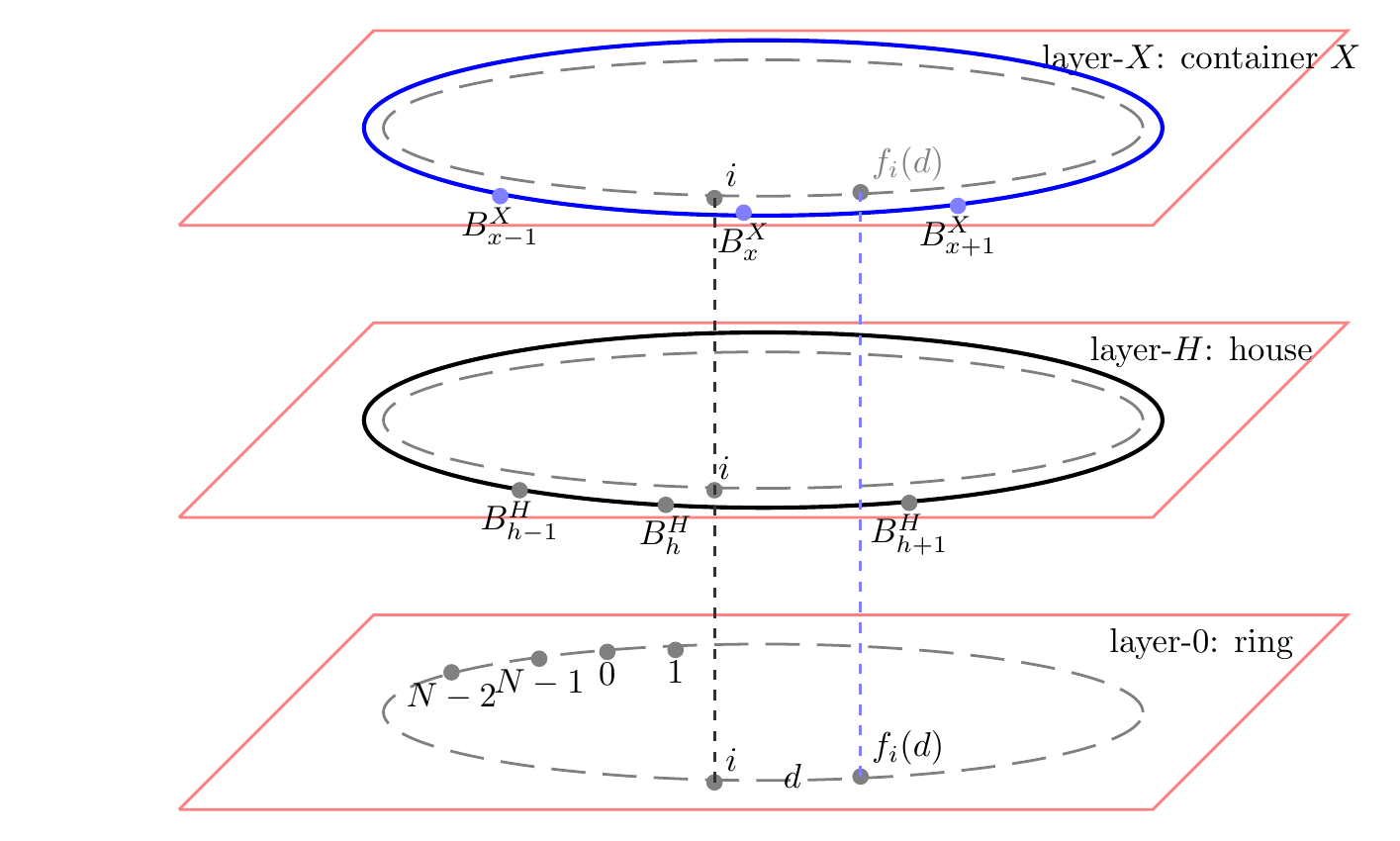}
	\caption{
		Multi-layer network scheme with vertices, location and layers of containers.
		Layer-0 is a 1D ring lattice $(N, k)$ with $k = 2$, 
		that represents the locations of vertices and is used to
		implement displacement and measure distance.
		At the house layer, displacement $d = 0$, therefore
		agent $i$ is assigned to house $h$ that covers its location,
		i.e., $B_{h}^{H} \le i < B_{h + 1}^{H}$.
		At other container layers-$X$, 
		displacement $d$ can be non-zero.
		Hence,
		while $B_{x-1}^{X} \le i < B_{x}^{X}$,
		$i$ is assigned to the container $x$ 
		since  $B_{x}^{X} \le f_{i}(d) < B_{x + 1}^{X}$.
	} 
	\label{fig:layers}
\end{figure}

\subsubsection{Locations}

In real-world, individuals' locations on earth can be represented by
$X$ coordinate, $Y$ coordinate and time information (ignoring the $Z$ axis).
Having an interest in individuals' location in a more broad sense
as in where they live in
rather than their exact location on each point in time,
temporal dimension can be omitted.
This leaves us with $X$ and $Y$ coordinates.

Inspired by earlier models,~\cite{
	watts1998collective%
},
this can be further simplified using
an auxiliary 1D ring lattice network,
where each vertex is of degree $k = 2$.
Then an individual is represented
by a vertex on the ring lattice with a
unique location denoted by its index.
See the ring lattice in layer 0 in \reffig{fig:layers}.

\subsubsection{Locality and Displacement}

People tend to work close to their home,
attend a nearby school, shop and have friends in the neighborhood.
This leads us to the \emph{locality},
which can be defined as
interactions taking place close to where people live.

In order to implement locality, 
we use a normal distribution with 0 mean and 
a 1D ring lattice.
Vertices are represented by indices in $\{ 0, 1, \dotsc, N-1 \}$ 
as in the 1D ring lattice in layer 0 in \reffig{fig:layers}.
We define the \emph{distance} between vertices $i$ and $j$ as the geodesic distance on 1D ring.
Consider agent $i$ in the 1D ring.
Starting at  $i$, 
if we move $d$ steps, 
which is called \emph{displacement}, 
the new location would be  $j = f_{i}(d)$,
where $f_{i}(d) = i + d$ in $\pmod{N}$.
Note also that for small values of $d$, the distance between $i$ and $j$ is small.
That is, $i$ and $j$ are local to each other.

We use this displacement to map an agent to a new location and associate 
he or she to 
(i)~the container that contains that location for the clique interactions;
(ii)~the agent at that location for the star case
as follows.

\subsubsection{Assignment to containers}
\label{sec:AssignmentToContainers}

Consider layer-$X$, such as blue-collar.
For agent $i$ on $X$,
assign $i$ to container $k$
if $B_{k}^{X} \le f_{i}(d) < B_{k + 1}^{X}$, 
where 
$B_{k}^{X}$ and $B_{k + 1}^{X}$ are the boundaries of the $k$th container
as in layer-0 and layer-$X$ in \reffig{fig:layers}.
If there are $N_{X}$ containers with capacities $\{ c_{k}^{X}\}_{k = 1}^{N_{X}}$ then the bounds can be calculated by
\begin{align*}
	B_{0}^{X} &= 0,\\
	B_{k}^{X}  &= B_{k - 1}^{X}
		+ \frac
			{c_{k}^{X}}
			{\sum_{\ell = 1}^{N_{X}} c_{\ell}^{X} }
			N
		\text{ for } k = 1, \dotsc, N_{X}
\end{align*}
as shown in layer-$X$ in \reffig{fig:layers}.

Setting the displacement $d = 0$ for house layer puts each agent into its home in \reffig{fig:layers}.

For other layers, displacement $d$ is sampled from a Gaussian distribution 
$\hN{\mu_{Xd}}{\sigma_{Xd}}$,
where $\mu_{Xd} $ is set to $0$ to satisfy locality.
Since $d$ is drawn from a Gaussian distribution,
$d$ can be a positive or negative real number.
In order to handle real values of $d$,
we need to refine our mapping with rounding as $f_{i}(d) = \textrm{round}( i + d )$ in $\pmod{N}$.

Note that every agent must have a home.
Therefore, the total capacity of houses is $N$.
Clearly not every agent must be in a container in other layers.
For example, 
an agent may be in school layer but not in blue-collar layer.
Hence, 
the total capacity of layers blue, white-collar, and school layers is strictly less than 
the total population.
That is,
we have $\sum_{k = 1}^{N_{X}} c_{k}^{X} \le N$.

\subsubsection{Assignment to star connections}
\label{sec:AssignmentToStarConnections}

Consider layer-$X$ such as friendship layer.
For agent $i$ on $X$, 
the number of connections $k_{i}$ is sampled from a Gaussian distribution 
$\hN{\mu_{X}}{\sigma_{X}}$.
For each connection agent $i$ is connected to some $j = f_{i}(d) $,
where displacement $d$ is, as usual, sampled from a Gaussian distribution 
$\hN{\mu_{Xd}}{\sigma_{Xd}}$.
Sampling $d$ is carried out for every connection separately.

See 
\reftbl{tbl:paramOthers},
\reftbl{tbl:paramLayers27} and
\refsec{sec:SettingParameters} 
for discussion of parameters 
$N_{X}$, 
$c_{k}^{X}$, 
$\mu_{Xd}$, 
$\sigma_{Xd}$,
$\mu_{X}$, and 
$\sigma_{X}$.

\begin{figure*}[!htp]
	\centering
		\begin{subfigure}[b]{\linewidth}
		\centering 
		\includegraphics[width=\columnwidth]%
			{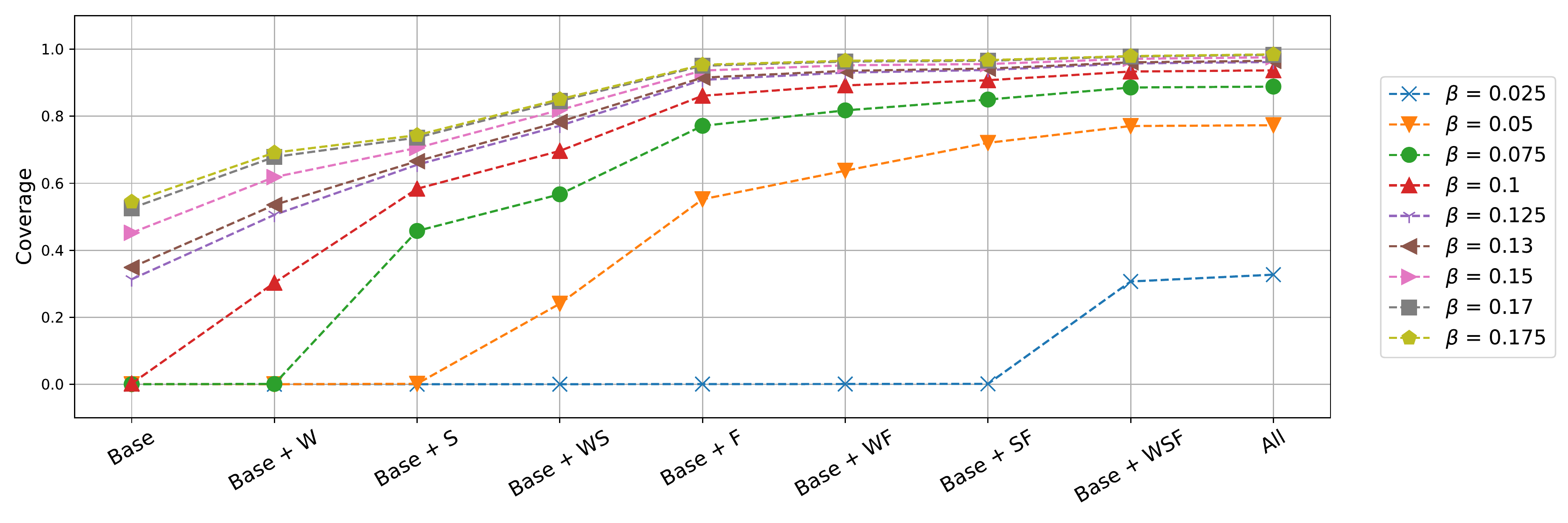}
		\caption{
		~(i) For real-life $\beta$ values of Covid-19,
			significant portion of the population is infected
			despite lockdowns.
		~(ii) Friendship is the most impactful layer to spread a disease.
		} 
		\label{fig:SIR_coverage}
		\end{subfigure}

	\hfill

		\begin{subfigure}[b]{\linewidth}
		\centering 
		\includegraphics[width=\columnwidth]%
			{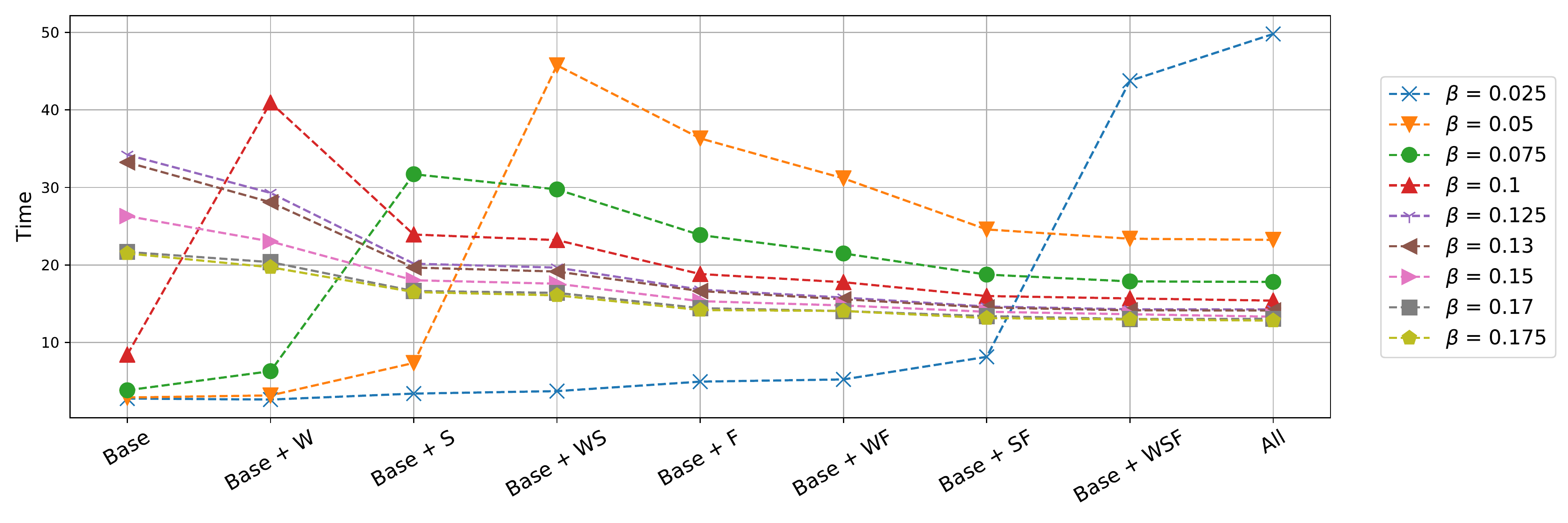}
		\caption{
		Once the disease is able to spread to above 0.20 coverage,
		additional layers cause a decrease in time, as more edges result in
		better disease spreading.} 
		\label{fig:SIR_time}
		\end{subfigure}
        \label{fig:SIR}
        \caption{
                Coverage and time as a function of various lockdown 	
                scenarios for a range of $\beta$ values. 
                $Y$-axes show coverage and time, respectively.
                \emph{Coverage} is defined as the portion of the population infected at some point in the simulation.
                \emph{Time} is defined as the time elapsed until the end of the simulation.
                $X$-axis shows different scenarios. 
                \texttt{Base} consists of  
                Household layer (L1), 
                Blue-collar work layer (L2), and 
                Service industry layer (L6), 
                since these three layers persisted throughout lockdown and curfews. 
                Layers \texttt{W}, \texttt{S}, and \texttt{F} stand for 
                white-collar (L3), 
                school (L4), and 
                friendship (L5) layers, respectively.
                Combinations of these, such as \texttt{Base+WF} indicate denoted layers are active simultaneously.
                \texttt{All} indicates the pre-Covid-19 world with no restrictions at all.
        }
\end{figure*}

\subsubsection{Role assignment}

Vertices are assigned to blue, white, and student groups randomly
according to their ratio in population, that is $\Gamma_{X}$.
For example, in a network
where $20~\%$ of the population goes to school,
each vertex has 0.20 chance to be labeled as a student.
This process is carried out for all containers and vertices.

In case the school layer is active, $T$ teachers are assigned to each class
from the nearest work container that contains at least $T$ number of
available employees who are not assigned to another class.

\subsection{Granularity of the model}

The consequence of such a model is that
every vertex is unique.
This depends on and is supplied by:
\begin{itemize}
	\item where the vertex resides and with whom he or she lives with
	\item whether he or she works/goes to school and connections he or she makes in work/school
	\item the number and the identity of his or her friends
	\item random connections he or she makes in his or her neighborhood.
\end{itemize}
Moreover, all of these connections occur locally, that is,
mainly in the vertex' neighborhood.
Hence, individuals are very different in terms of 
where they live, whom they know in different networks
(work, school, neither),
and to whom they can transmit disease.

\subsection{SIR}

Network connectivity depends on the choice of layers.
We remove the layers that we want to lock down.
Note that the house and blue-collar layers are not sufficient to 
obtain a connected network.
Therefore, 
disease stays in the connected component, 
which contains the initial infected vertex.
That is, 
it cannot reach the entire network.
Additional layers begin to connect the network.

Having a network that is ready to be inspected,
we conduct agent-based SIR simulations, 
starting from a single infected vertex.
At each timestep an infected vertex $i$ infects another vertex $j$
with probability $\beta$, given they are connected by
an edge with a weight of $\beta$.
At the same time, infected vertices become recovered with
recovery rate $\gamma$.
This is repeated until there is no infected vertex in the network,
and the simulation ends.
Then we record the coverage and the time.
\emph{Coverage} is defined as the ratio of agents that receive the infection,
and \emph{Time} is how long it takes until the simulation ends.

Coverage depends on the initial agent.
To account for the worst-case scenario,
we consider the agent
with the highest \emph{strength}~\cite{barrat2004strength}, 
that is, the sum of edge weights of a vertex, 
from the innermost core~\cite{%
	dorogovtsev2006kcore,%
	kitsak2010identification,%
	atdag2021computational}
of the largest component of the network.
In this way, we look for the worst case in the given scenario and stabilize the potential
high variance in simulation results
that otherwise could be caused by random choice of initial infected vertex.
We use \texttt{fast\_SIR} simulation from Epidemics on Networks EoN package~\cite{%
	miller2019eon,%
	miller2020eon},
and set recovery rate $\gamma = 1$ for all experiments.

In order not to be specific to a network, 
which is created by many stochastic processes such as 
random number generation and sampling from different distributions,
we create a new network in each run we take.
Therefore, in each run, we create a network with selected parameters, find the best spreader vertex in the largest component and 
start the SIR simulation by infecting this vertex.

\begin{table*}[htp]
\centering
\caption{
	Network Attributes.
}
\begin{tabular}{|l r r r r r r r|}
	\hline
	Network Attribute& [L1] & [L1-L2] & [L1-L3] & [L1-L4] & [L1-L5] & [L1-L6] & [L1-L7] \\ 
	\hline
	\hline
	Size of largest component & 0~\%& 47~\% & 83~\% & 96~\% & 100~\% & 100~\% & 100~\%\\
	Diameter & 1.00 & 38.44 & 24.78 & 18.71 & 9.43 & 7.45 & 6.77\\
	Average shortest path length & 1.00 & 16.65 & 11.32 & 8.31 & 4.81 & 4.14 & 3.61\\
	Average clustering coefficient & 0.66 & 0.65 & 0.65 & 0.70 & 0.18 & 0.12 & 0.05\\
	\hline
\end{tabular}
\label{tbl:NetworkAttributes}
\end{table*}

\section{Experiments and Observations}

Median of coverage and time of 300 realizations 
in this setting are shown in \reffig{fig:SIR_coverage}.
We examine different scenarios starting with \texttt{Base}, 
which consists of layers L1, L2, and L6.
We consider this as a baseline scenario
since these three layers were the most fundamental layers,
persisting even in times of lockdown and curfews for the survival of society.

Then we continue by adding one layer at a time, like (\texttt{Base+W}),
where we send white-collar agents to work. 
The combination of multiple letters followed by \texttt{Base} indicates that 
layers corresponding to these letters
were active simultaneously in that scenario.
For example, 
(\texttt{Base+WS}) means white-collars go to work, 
schools are open with students and teachers going to classes physically,
but curfews still existing with no socialization with friends or neighbors, and no traveling.

\subsection{Coverage}

We start with reporting coverage,
the ratio of infected vertices over all vertices,
in \reffig{fig:SIR_coverage}.
As expected, 
for low values of $\beta$,
\texttt{Base} network by itself is not enough to obtain disease spread.
\reffig{fig:SIR_coverage} indicates that 
we need all layers (\texttt{All}) in order to reach a nonzero coverage for  $\beta = 0.025$.
We need to increase $\beta$ value to $0.125$ 
in order to get nonzero coverage for \texttt{Base} layers only.
If we consider adding one single layer to the \texttt{Base},
friendship is the first layer to produce nonzero spread at $\beta = 0.05$.
At  a higher value of $\beta = 0.075$, the \texttt{Base} and school pair (\texttt{Base+S}) follows.
Then comes \texttt{Base} and white-collar (\texttt{Base+W}) layers.
In fact, the \texttt{Base} and friendship combination (\texttt{Base+F}) provides the highest coverage 
compared to all other pairs of single layer on top of \texttt{Base}, for $\beta > 0.025$.
Considering \texttt{Base} and two other layer combinations, 
\texttt{Base}, friendship, and school (\texttt{Base+SF}) combination has the highest coverage.

Considering the reported Covid-19 transmission rates 
$\beta = 0.13$~\cite{zhang2021beta013} and 
$\beta = 0.17$~\cite{feng2021beta017},
important observations of \reffig{fig:SIR_coverage} are:
\begin{itemize}
	\item 
	Disease is able to reach a significant portion ($\approxeq 40$\%) of the population
	despite complete lockdown \texttt{Base}
	when $\beta > 0.1$.
	
	\item 
	True outbreak with coverage larger than $0.8$ occurs 
	in scenarios that include friendship layer 
	when $\beta > 0.1$.
	
	\item 
	Friendship layer (\texttt{Base+F}) is the single most impactful layer, and
	even combined white-collar and school layers (\texttt{Base+WS}) are not as effective at spreading disease.
	
	\item 
	Remote work (\texttt{Base+SF}) is not very effective in slowing down 
	disease compared to remote school (\texttt{Base+WF})
	or restricting socialization with friends (\texttt{Base+WS}).

	\item 
	Majority of the population is infected for all scenarios except \texttt{Base},
	with reported Covid-19 transmission rates
	$\beta = 0.13$ and $\beta = 0.17$.
	The spread reaches almost the entire population when friendship layer is active.
	
	\item
	For high values of $\beta > 0.125$,
	we observe a saturation above $0.8$ coverage for \texttt{Base} with any two or more layers.
\end{itemize}

\subsection{Time}

Then we inspect the time in \reffig{fig:SIR_time}
to understand how long it takes to obtain
the corresponding coverage values.
We observe two distinct patterns.
\begin{itemize}

	\item
	For $\beta$ values larger than 0.10, 
	addition of each layer results in a reduction in time.
	This is because each additional edge results in a
	network that is more connected than before.
	As a result, as observed in \reftbl{tbl:NetworkAttributes},
	diameter and average shortest path length of the network
	decrease.
	Consequently, the disease spreads to the rest of the network faster,
	and the simulation ends more quickly.
	
	\item
	For $\beta$ values less than 0.125, time first increases, then decreases. 
	What is different here is the initial increase.
	It happens because the disease is able to spread to
	significantly more vertices
	for the first time with the introduction of the corresponding layer(s).
	Looking at the corresponding coverage values in
	\reffig{fig:SIR_coverage}, we observe that the disease was unable to
	infect more than 0.20 of the population prior to that.
	Hence, the reason the time increases with additional layer(s)
	is because the disease spreads to a larger portion
	of the population, and this takes time.
	Once it reaches above 0.20 coverage, 
	additional layers allow faster spread and lower time values.
\end{itemize}

These two patterns are observed with a shift in lines
for different $\beta$ values.
This is expected behavior since higher beta values
result in faster disease spreading.

\subsection{Network Attributes}
\label{sec:NetworkAttributes}

In this section, network attributes are inspected as layers are added up
and are reported in \reftbl{tbl:NetworkAttributes}.
The leftmost column indicates the inspected attribute and other columns indicate
the layers that are present,
such as [L1-L3], 
which means the first three layers are active
and the other four layers are not.
Reported numbers are the average values for 300 networks that are generated
with the same exact parameters, 
which are discussed in detail in \refsec{sec:SettingParameters}.
In case the network is not connected,
the attribute is measured for the largest component of the network.

In the case of [L1], the network consists of disconnected components,
hence size of the largest component as \% is approximately 0,
and the diameter and average shortest path length
for the largest component are 1.
It is observed that [L1-L3] is sufficient to connect 83~\% of the vertices.
Further layers improve the connectedness of the network,
increasing the size of the largest component and decreasing diameter
and average shortest path length.

The network exhibits strong clustering structure
for four phases of layers, starting at [L1] and ending at [L1-L4].
The addition of L5 causes a significant drop.
This behavior is expected since L5 is the first layer
that is not container type, and it connects a vertex to
several vertices that are part of several containers.

We believe that it is more plausible to take [L1-L5] into account
when comparing the proposed model's network attributes
to real-life network attributes
rather than [L1-L7] or [L1-L6]
since L6 and L7 represent exponentially weaker relations
as explained in ~\refsec{subsec:ServiceIndustry},
that are ignored in real-life networks.
Herewith, inspecting the properties of [L1-L5],
it is observed that our model successfully generates networks
that exhibit similar properties
to real-life networks, 
namely small diameter, and average shortest path length and
high clustering coefficient~\cite{
	newman2006structure,%
	krioukov2010hyperbolic,%
	zuev2015emergence%
}.

\subsection{Role of other parameters}

In \reftbl{tbl:paramLayers27}, it is observed that
most of the parameters are empirical,
that is, they come from real-life statistics.
The rest are plausibly assumed values.
In order to see the impact of assumed parameters,
we conduct additional experiments
where we examine the roles of
(i)~$\mu$ of layers 6 and 7,
(ii)~$\sigma$ of layers 2 to 7, and
(iii)~$\sigma_{0}$.
We have conducted our experiments by
(i)~increasing and
(ii)~decreasing
the values in 
\reftbl{tbl:paramLayers27},
by 50\%.
Then we run the same SIR simulation to see
their impact on coverage and time.
It is observed that
\begin{itemize}
	\item
	Increasing $\mu$ and $\sigma$ for layers 6 and 7 allows
	the disease to spread better, and vice versa.

	\item
	Increasing $\sigma$ for layers 2 to 5 allows
	the disease to spread better as well.
	However, its impact is more significant compared to
	the modification of $\mu$ and $\sigma$ for layers 6 and 7.
	This is due to two reasons.
	First, an increase in connectivity
	of four layers makes more impact than two layers.
	Second, $\beta$ values of the first five layers are significantly
	larger than the last two layers.

	\item
	Similarly, increasing $\sigma_{0}$ allows the disease
	to spread better, and vice versa.
	This is because connections to further corners of the network
	results in a smaller average shortest path and better connectivity.
	
\end{itemize}

Overall, it is intuitive that more edges
to distant vertices allow better transmission.
However, we find that the effect of these parameters are relatively
insignificant as they do not impact the results for real-life
$\beta$ values.
Moreover, the characteristic of the figures was not affected by
these parameters, that is, the relative impact of each layer.
For this reason, we do not report additional figures
for these experiments.
Conclusively, change in these parameters do not change
the main findings in this work.

\section{Discussion}

Evidential results from real-world data show that
the top two most effective non-pharmaceutical interventions (NPI) against Covid-19 are 
small gathering cancellation and closure of educational institutions~\cite{haug2020nature}.
This is consistent with the results of our model,
where the most important layer is socialization with friendship layer,
followed by the school layer.
The high impact of friendship layer arises from two reasons.
First, it is an intimate relation with high a transmission rate.
Second, and more importantly,
it connects clusters of house and work containers
that are otherwise disconnected or weakly connected.
Both concepts are explained in detail in \refsec{sec:NetworkGenerator}.

Following that, Gosak's recent work~\cite{gosak2021community} 
onto the effectiveness of community lockdowns indicates that lockdowns are
effective only if communities in the network are disjointed.
This is aligned with \reffig{fig:SIR_coverage} where disease is able to spread to
about half of the network for reported Covid-19 $\beta$ values, 0.13 and 0.17,
even in the case of \texttt{Base}.

\subsection{Limitations}

\textbf{Predecessor-successor edges.}
Our model does not take time into account in terms of predecessor-successor edges.
Suppose a susceptible vertex $i$ contacts another susceptible vertex $j$.
Then $i$ contacts the infected vertex $k$, and gets infected.
In this scenario, $i$ is infected after contacting $k$,
therefore, he or she cannot infect $j$ 
because $i$ was not infected back then.
Our framework does not model this type of time dimension when creating edges.

\textbf{Gaussian distribution.}
We use a set of parameters,
some of which define distributions that are used
throughout network creation process.
If a distribution is known, we use it, 
as in the case of household size distribution,
which is right-skewed Gaussian distribution~\cite{lfs2017turkey}.
If it is not known, 
we assume it is Gaussian.

\textbf{Locality.}
In our model, most of the interactions prefer locality,
that is, an interaction between two distant vertices is unlikely
compared to an interaction within the neighborhood.
Therefore, we assume that all displacement measures come from
Gaussian distribution with $\mu_{d} = 0$ and $\sigma_{d} = 1000$
for all layers.
We have no information about how strong locality is for different layers in real-life.

\textbf{Exponentials of $\beta$.}
We assume different types of interactions have different $\beta$ transmission probabilities
and simplify this by using exponentials of $\beta$ in different layers.
In this way, $\beta$ decays rapidly from intimate relations to short-duration ones.
This is plausible when comparing a contact of 8 hours with one of 30 seconds,
but it is still an assumption.

\subsection{Future Work}

\textbf{Diffusible spread on networks.}
Even though the experimental focus in this work 
has been onto Covid-19 due to the recent pandemic,
proposed model can be used to inspect the spread on networks
of virtually any diffusible such as disease, gossip, idea.
Parametric weighted network structure especially facilitates
research on disease spread, 
e.g., future variants of Covid-19 with corresponding transmission rates.

\textbf{Vertex assortativity.}
We assign vertices to houses and create friendship connections randomly,
but it may be more realistic to consider assortativity~\cite{newman2002assortativity} 
when building these relations
as it may be more likely that similar vertices will live together and befriend each other
as a result of socioeconomic and demographic factors.

\textbf{Multiple initial infected agents.}
Our model starts with all agents in susceptible state except one infected.
We try to select the infected one among the agents with the largest spread capacity.
The study of initial multiple infected vertices is left for future work.

\textbf{Multiple towns.}
This model is designed to inspect single town scenarios in high resolution.
To represent a larger scale real-world,
multiple towns can be generated with an additional layer where
vertices of a town will make connections with vertices in another town
to represent travel.
Such a model can help understand the transmission of disease
between towns and countries~\cite{davis2021cryptic}.

\textbf{Advanced variants of SIR.}
In this work, we used the simple SIR model.
More sophisticated variants such as SIRS, SEIR and MSIR
can be run on our model for more detailed and realistic
analysis of disease spreading.

\section{Conclusion}

Witnessing the absence of a high resolution network generator in literature,
we offer a parametric multi-layer undirected weighted network scheme
to model a hypothetical urban town where
individuals and their interactions are represented as vertices
and weighted edges, respectively.
Multi-layer networks are utilized to represent various interactions
with different transmission rates, each layer corresponding to
a different fundamental relation in everyday life.
The layered architecture makes it possible
to lock down different combinations of layers
to model scenarios like remote work, remote school, and curfews.
First, we run SIR simulations on generated networks
for different lockdown scenarios and show that the friendship layer
is the most impactful layer to slow down epidemics.
Second, we inspect and compare our model's generated networks' attributes
to real-life networks.
It is observed that our model's attributes and simulation results
are aligned with real-life data and the most recent research.
This indicates the strength and realism of our network generator model
and stimulates network science research.

\section*{Acknowledgment}

We would like to thank Erol Taymaz, Suzan Uskudarli, Emre Aladag and Samet Atdag 
for constructive comments.
This work is partially supported by 
the Turkish Directorate of Strategy and Budget
under the TAM Project number 2007K12-873.

\begin{appendices}

\section{Setting parameters}
\label{sec:SettingParameters}

Network generation requires a number of parameters.
Starting by creating a network with $N_{H} = 10,000$ houses, 
which corresponds to a network of approximately $N = 26,400$ vertices,
we use statistics from the US whenever available
and assume plausible values for those that are not.
Arbitrary choice of $N_{H} = 10,000$ is due to the feasibility of computation,
hence larger networks can be created with an additional cost of computation and memory.
Collected and assumed parameters are shown in \reftbl{tbl:paramOthers} and \reftbl{tbl:paramLayers27}.

\begin{table*}[htp]
\centering
\caption{
	Parameters defining layer 1.
}
\begin{tabular}{|c c l l|} 
	\hline
	Parameter & Value & Description &~\\ 
	\hline
	\hline
	$N_{H}$ & 10,000 & Number of houses & \\ 
	$\alpha$ & 3.96 & Shape of household size skewnorm distribution  & \cite{lfs2017turkey, statista2021muhouse}\\
	$\xi$ & 1.22 & Location of household size skewnorm distribution  & \cite{lfs2017turkey, statista2021muhouse}\\
	$\omega$& 1.75 & Scale of household size skewnorm distribution & \cite{lfs2017turkey, statista2021muhouse}\\
	$T$ & 3 & Number of teachers assigned per class  & \\%
	\hline
\end{tabular}
\label{tbl:paramOthers}
\end{table*}

\begin{table*}[htp]
\centering
\caption{
	Parameters defining layers 2-7, 
	where 
	$\mu_{0} = 0$ and $\sigma_{0} = 1000$.
}
\begin{tabular}{|	ll	| rl	| rlr	| ll	|	r|}
	\hline
	Layer	 &$X$	 &$\Gamma_{X}$	 &	 &$\mu_{X}$	 &	 &$\sigma_{X}$	 &$\mu_{Xd}$	 &$\sigma_{Xd}$	 &$\beta_{X}$ \\
	\hline
	\hline
	2: Blue workforce	 &$B$	 &21.0~\%	 &\cite{statista2021workforce, worldometers2021uspopulation, usnews2021ratiowhite}	 &7.6	 &\cite{gallup2020muwork,del2007muwork,theladders2018muwork}\	 &3	 &$\mu_{0}$	 &$\sigma_{0}$	 &$\beta^{1}$ \\
	3: White workforce	 &$W$	 &27.0~\%	 &\cite{statista2021workforce, worldometers2021uspopulation, usnews2021ratiowhite}	 &7.6	 &\cite{gallup2020muwork,del2007muwork,theladders2018muwork}\	 &3	 &$\mu_{0}$	 &$\sigma_{0}$	 &$\beta^{1}$ \\
	4: Students	 &$S$	 &24.7~\%	 &\cite{census2018ratioschool}\	 &19.6	 &\cite{nces2018muschool}	 &3	 &$\mu_{0}$	 &$\sigma_{0}$	 &$\beta^{1}$ \\
	\hline
	5: Friendship	 &$F$	 &- &- 	 &12.3	 &\cite{independent2019mufriends,gallup2004mufriends}\	 &5	 &$\mu_{0}$	 &$\sigma_{0}$	 &$\beta^{1}$ \\
	6: Service industry	 &$C$	 & 15.0~\%	  &\cite{cpsaat2020ratioserviceindustry}	 &50.0	 &	 &20	 &$\mu_{0}$	 &$\sigma_{0}$	 &$\beta^{2}$ \\
	7: Random encounters	 &$R$	 &-	 &-	 &50.0	 &	 &20	 &$\mu_{0}$	 &$\sigma_{0}$	 &$\beta^{3}$ \\
	\hline
\end{tabular}
\label{tbl:paramLayers27}
\end{table*}

\subsection{Layer~1. Household}

According to ref~\cite{lfs2017turkey},
the average number of households in Turkey is 2.53 
with a skewed normal distribution,
which is defined by $f(\alpha = 3.96, \xi = 1.22, \omega =  1.75)$~\cite{azzalini1999skewnorm} 
with parameters shape, location, and scale, respectively.
According to ref~\cite{statista2021muhouse}, 
the average household size for the US in 2020 is 2.53.
Since we do not know the true distribution of household size for the US
but expect it to have very similar characteristics to the distribution for Turkey,
which has the same mean,
household size is determined by sampling from this distribution.

Since household connections are the most intimate with
the highest transmission probability,
we assume an infected vertex will surely infect others in its home,
therefore we set $\beta_{1} = \beta^{0} = 1$.

\subsection{Layers~2-3. Work}

In this work, differentiation between blue and white-collar layers exists solely 
to be able to modularly model
employees who work from home during a lockdown.
Hence the only difference between blue and white-collar layers is their ratio in population, 
$\Gamma_{W}$ and $\Gamma_{B}$, and other parameters are the same for both groups.

According to references~\cite{%
	gallup2020muwork,%
	del2007muwork,%
	theladders2018muwork%
}, 
the number of people interact within a workplace are 9.8, 8 and 5, respectively.
We use the mean of these three values, 7.6, as our $\mu_{W}$ and $\mu_B$ parameters,
and assume $\sigma_{W}$ and $\sigma_{B}$ to be 3.

Prior to Covid-19,
$48~\%$ of the population was in the workforce in the US~\cite{%
	statista2021workforce,
	worldometers2021uspopulation}.
This ratio is our baseline when creating jobs and employees.
As of January 2021,
$56~\%$ of the workforce worked remotely~\cite{usnews2021ratiowhite}.
Using these two data,
we obtain the ratios 
$\Gamma_{W} = 0.48 \cdot 0.56 = 27~\%$
and
$\Gamma_{B} = 0.48 - 0.27 = 21~\%$ for 
white-collars and blue-collars, respectively.
Hence, we create workplaces and vertices of white and blue according to these parameters.

Work relations are not as intimate as households, but employees still spend several hours a day together,
thus we set $\beta_{2} = \beta_{3} = \beta^{1}$.

\subsection{Layer~4. School}

Even though a school consists of several classrooms 
where students may also interact and play with students outside the classroom,
this is a rather weaker and less likely relation compared to in-class relations,
so it is neglected for simplicity and only the interactions 
in-classroom are modeled in this work.

Ref~\cite{census2018ratioschool} indicates that
$\Gamma_{S} = 24.7~\%$ of the population was enrolled in schools nationwide in 2017.
Ref~\cite{nces2018muschool} provides the average class size for states in the US.
Taking the mean across this sheet for both axes, we obtain $\mu_{S} = 19.6$.
Having no information about this distribution, we assume $\sigma_{S} = 3$.
Although the number of teachers in a classroom depends on the education level and other factors, 
we simplify this to  $T = 3$.

School relations are very similar to work relations 
in terms of duration and being in containers,
so we set $\beta_{4} = \beta^{1}$, as well.

\subsection{Layer~5. Friendship}

The average number of friends a person has varies according to different sources~\cite{%
	gallup2004mufriends,%
	independent2019mufriends%
},
being 8.6 and 16, respectively.
We choose the average of the two and
set $\mu_{F} = 12.3$, and assume $\sigma_{F} = 5$, 
which allows both small and large number of friends for different vertices.

Assuming that the friendship relation is at least as intimate as work or school layer,
we set $\beta_{5} = \beta^{1}$.

\subsection{Layer~6. Service industry}
\label{subsec:ServiceIndustry}

In addition to the first and second layers, one last layer persisted throughout lockdown,
virtually everyone still needing essential services such as foods, logistics, health care.
Consequently, potentially everyone made connections with workers in these businesses, 
such as cashiers and couriers.
In fact, workers of these essential services were in contact with many people a day.
The ratio of service industry workers in population is denoted by $\Gamma_{C}$.

The US Bureau of Labor Statistics provides detailed figures on the US 
in terms of headcount and demographics for each sector in detail~\cite{cpsaat2020ratioserviceindustry}.
According to our definition, which is trivially a subset of blue-collar workers,
the service industry consists of 
`Wholesale and retail trade`, 
`Taxi and limousine service`, 
`Couriers and messengers`,  
`Real estate and rental and leasing`, 
`Veterinary services`,
`Services to buildings and dwellings`, 
`Health care and social assistance`, 
`Accommodation and food services`, 
`Other services, except private households` 
elements in the ``cpsaat2020'' table.
The total number of people employed in these services 
divided by the total workforce corresponds to $20~\%$ of the population.
However, this is not very accurate for two reasons: 
First, $\Gamma_{B} = 21~\%$ already, and
blue-collar work is not almost entirely made of service industry. 
Second, not all employees in these sectors are in fact blue-collar workers.
Therefore, to make it more realistic and plausible, 
we multiply this $20~\%$ by a coefficient of 
$\frac{3}{4}$ and obtain $\Gamma_{C} = 15~\%$, 
which defines the number of employees in the service industry 
who are in active contact with customers.

Since we have no statistical data on how many contacts a service industry worker
makes in a given time interval,
we assume $\mu_{C} = 50$ and $\sigma_{C} = 20$,
which has the ability to represent a wide range of jobs.

Compared to other relations, contact between
the service provider and customer lasts much shorter.
Therefore we set  $\beta_{6} = \beta^{2}$, 
which results in an exponentially lower transmission probability than earlier layers.

\subsection{Layer~7. Random Encounters}

Interactions people make in daily life do not consist of relations between
households, colleagues, students in class, friends known,
or cashiers in local stores only.
Random encounters with unknown people occur daily during
shopping, traveling, or
simply walking by another person.

We also have no prior information about the number of random encounters, 
so we assume
$\mu_{R} = 50$ and $\sigma_{R} = 20$.

We believe random encounters have even a shorter duration
with lower transmission probability 
compared to six layers defined so far.
Thus, we set $\beta_{7} = \beta^{3}$ with an even lower transmission probability.

\subsection{Locality}

We assume that displacement $d$ for locality comes from a Gaussian distribution $\hN{\mu_{0}}{\sigma_{0}}$.
We set $\mu_{0} = 0$ so that displacement can be either positive or negative.
We assume that $\sigma_{0} = 1000$ for all layers 2-7.

\subsection*{Code availability}
The source code of the proposed network generator can be accessed at 
\url{https://github.com/meliksahturker/NetGen}.

\end{appendices}

\bibliographystyle{IEEEtran}
\bibliography{pCovid-v13}

\end{document}